\documentclass[twocolumn,prc,aps,floatfix]{revtex4}
\usepackage[dvips]{graphicx}
\begin{document}
\title{Universal scaling of Efimov resonance positions in cold atom systems}
\author{Cheng Chin}
\affiliation{James Franck Institute and Physics Department, The
University of Chicago, IL 60637}

\date{\today}

\begin{abstract}
Recent cold atom experiments report a surprising universal scaling of the first Efimov resonance position $a_{-}^{(1)}$ by the two-body van der Waals length $r_{vdW}$. The ratio $C=-a_{1}^{(1)}/r_{vdW}=8.5\sim9.5$ for identical particles appears to be a constant regardless of the atomic spin configuration, the Feshbach resonance employed to tune the scattering length, and even the atomic species, with $^{39}$K being the only exception. This result indicates that the Efimov energy structure is insensitive to the details of the short range potential. We suggest that the universality results from the quantum reflection of the Efimov wavefunciton by the short-range molecular potential. Assuming Born-Oppenheimer approximation and strong quantum reflection, we obtain an analytic result of $C=9.48...$ for three identical particles. We suspect the exceptional case of $^{39}$K is a result of resonant coupling between the Efimov state and a short-range molecular state.
\end{abstract}

\maketitle


Vitaly N. Efimov predicted in 1970 that a unique set of three-body bound states emerges when the interactions between any two of the three particles are resonantly enhanced \cite{Efimov1970ela}. While Efimov physics was discussed in the context of nuclear physics, they have attracted much attention in recent studies on collisions of cold atoms and molecules. Cold atoms are excellent candidates to explore Efimov physics since the binary atomic interactions can be controlled via a Feshbach resonance \cite{Chin2010fri}. When the two-body scattering length $a$ is tuned to a large and negative value, the emergence of an Efimov state near the atomic scattering continuum can be identified from the strongly enhanced recombination loss of trapped atoms \cite{Esry1999rot}. This resonance is also called an Efimov resonance. In principle, an infinite number of Efimov resonances are expected to appear when $a$ approaches (negative) infinity. The position of the $N-$th Efimov resonance at $a=a_{-}^{(N)}$ is predicted to follow a scaling law $a_{-}^{(N)}=c a_{-}^{(N-1)}$, where $c=e^{\pi/s_0}$ and $s_0$ are constants. For three identical bosons, one obtains $s_0\approx1.00624$ and $c\approx22.68$ \cite{Efimov1970ela, Efimov1979lep}.

The appearance of Efimov states and the recursion relationship can be understood as a result of an effective potential term in the scattering of three particles. For three identical particles, the Efimov potential is given, in the hyperspherical coordinate, by \cite{Efimov1970ela}

\begin{equation}
V_{E}(R)=-\frac{\hbar^2}{2m}\frac{s_0^2+1/4}{R^2} \,\,\,\,\,\,\,\,\,\mbox{(for $R_0<R<|a|$)},
\label{eq1}
\end{equation}

\begin{figure}\includegraphics[width=3 in]{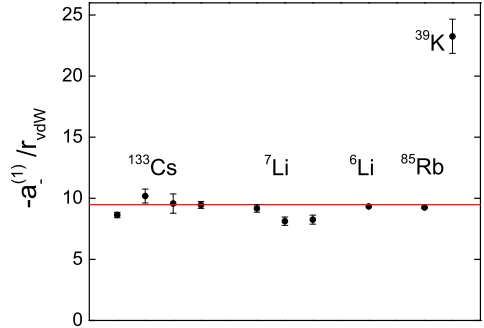}\\
\caption{Positions of the Efimov resonance reported in cold atoms. By tuning the two-body scattering length via Feshbach resonances, Efimov resonances are found in $^{133}$Cs \cite{Kraemer2006efe, Berninger2011uot},  $^{7}$Li \cite{Pollack2009uit, Gross2009oou}, $^6$Li \cite{Huckans2009, Ottenstein2008}, $^{39}$K \cite{Zaccanti2009ooa} and $^{85}$Rb \cite{Rb}. See Table~\ref{table} for details. The ratio of the resonance position to the two-body van der Waals length $C=-a_{-}^{(1)}/r_{vdW}$ is presented and compared to our prediction of $C=9.48...$ (solid line).}
\label{fig1}
\end{figure}

\noindent where $m$ is the atomic mass, $2\pi \hbar$ is Planck's constant, hyperspherical radius $R=(r_{12}^2/3+r_{23}^2/3+r_{31}^2/3)^{1/2}$ characterizes the size of the system, $r_{ij}$ is the distance between the $i-$th and $j-$th particle, and $R_0$ is the interaction range, assumed to be very small compared to $|a|$. The logarithmic divergence of the number of bound states is a direct consequence of the inverse square law of the Efimov potential $V_E\sim -R^{-2}$ when $|a|$ approaches infinity. For a finite $a$, the Efimov potential terminates at $R\approx |a|$ and a repulsive potential $\sim R^{-2}$ dominates for $R>|a|$ \cite{Incao2005}.

The recursion relation $a_{-}^{(N)}=c a_{-}^{(N-1)}$ also suggests that the full Efimov resonance spectrum can be determined when the lowest resonance position $a_{-}^{(1)}$ is known. The exact value of $a_-^{(1)}$, however, is expected to be non-universal since it depends on the three-body scattering phase shift $\phi$, which contains the details of the short range molecular potential \cite{Esry1999rot,Braaten2006uif, D'Incao2009}. The first resonance position $a_-^{(1)}$ can in principle assume any value between $R_0$ and $c\times R_0$.

\begin{table}
\caption{Efimov resonance positions, two-body van der Waals length $r_{vdW}$ and their ratios for several atomic species. Values of the van der Waals coefficient $C_6$ and $r_{vdW}$ are based on Ref.~\cite{Chin2010fri} and mass scaling. (1 Bohr radius $a_B$= 0.0529... nm and 1 a.u.= 1 $E_h a_B^6$ where $E_h$ is a hartree.) } \label{table}
\begin{tabular}{clccc}\hline
  Atom       & $a_{-}^{(1)}$ ($a_B$) & $C_6$ (a.u.)    & $r_{vdW}$ ($a_B$)  & $-a_{-}^{(1)}/r_{vdW}$ \\ \hline
  $^6$Li     & -292  \cite{Wenz2009uti}       & 1393.4          & 31.26             & 9.34  \\
  $^7$Li     & -298(10) \cite{Pollack2009uit} & 1393.4          & 32.49             & 9.17(31) \\
  $^7$Li     & -264(11) \cite{Gross2009oou}   & 1393.4          & 32.49             & 8.13(34) \\
  $^7$Li     & -268(12) \cite{Gross2010nsi}   & 1393.4          & 32.49             & 8.25(37) \\
  $^{39}$K   & -1500(90) \cite{Zaccanti2009ooa} & 3897            & 64.49             & 23.3(1.4) \\
  $^{85}$Rb  & -759(6) \cite{Rb}                & 4698       & 82.10             & 9.24(07)  \\
  $^{133}$Cs & -872(22) \cite{Berninger2011uot} & 6860 & 101 & 8.63(22) \\
  $^{133}$Cs & -1029(58) \cite{Berninger2011uot}                & 6860 & 101 & 10.19(57) \\
  $^{133}$Cs & -957(80)  \cite{Berninger2011uot}                & 6860 & 101 & 9.48(79) \\
  $^{133}$Cs & -955(28)  \cite{Berninger2011uot}                & 6860 & 101 & 9.46(28) \\ \hline
\end{tabular}
\end{table}

The first Efimov resonance was observed in cold collisions of bosonic cesium atoms \cite{Kraemer2006efe}. In this experiment, recombination loss of trapped cesium atoms shows a pronounced peak when the scattering length is tuned to $a=a_{-}^{(1)}=-850$~$a_B$, where $a_B$ is the Bohr radius. Following this work, recent experiments have observed Efimov resonances in the cold collisions of almost all alkali atoms. Figure~\ref{fig1} and Table~\ref{table} summarize the measurements based on single atomic species. A major surprise revealed in these measurements is the constancy of the ratio of the resonance position to the two-body van der Waals length $C=-a_{-}^{(1)}/r_{vdW}=8\sim10$, regardless of the atomic spin configuration \cite{Gross2009oou,Gross2010nsi}, the Feshbach resonances employed to tune the scattering length \cite{Berninger2011uot}, and even the atomic species (with $^{39}$K being the only exception). This result is unexpected and challenges the theoretical understanding of Efimov physics in cold atoms.

Theoretically, universality of Efimov physics across different Feshbach resonances in the same scattering channel was discussed in Ref.~\cite{Gemelke2008}. When the magnetic field is tuned to different Feshbach resonances, the field only induces coupling to different molecular states, and does not substantially modify the form nor the strength of the potential in the open channel. A fractional change of the potential less than $10^{-9}$ per Gauss is estimated for cesium atoms, which suggests the non-universal effects can remain unchanged when the field is tuned to different Feshbach resonances \cite{Gemelke2008}. However, this argument does not explain the universality for different spin configurations or atomic species.

The observed broad universality, as shown in Fig.~\ref{fig1}, indicates that Efimov physics in cold atoms can in principle be captured by a general physics picture that does not depend on the details of the short range molecular potential, Zeeman shift, and spin configuration... The goal of this paper is to present one such model and compare our prediction with the measurements.

\begin{figure}\includegraphics[width=3.5 in]{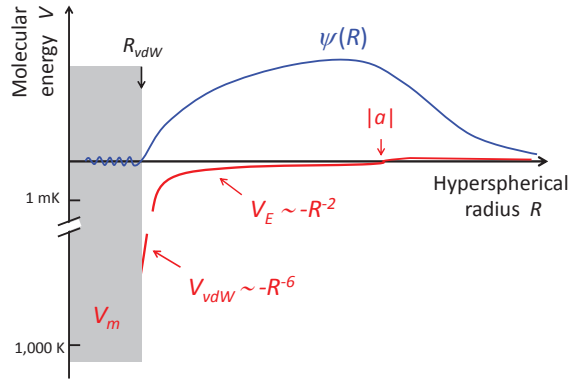}\\
\caption{Illustration of three-atom interaction potential and the wavefunction of the first Efimov state. In the hyperspherical coordinate, Efimov potential $V_E(R)=-(s_0^2+1/4)\hbar^2/(2mR^2)$ dominates for $R_{vdW}<R<|a|$, and is replaced by a repulsive potential for $R>|a|$ \cite{Incao2005}. Near $R=R_{vdW}$, van der Waals potential $\sim -1/R^6$ quickly takes over the Efimov potential. At even shorter distances, molecular potential $V_m$ (shaded area) becomes complex and deep, reaching  $10^3\sim10^4$K in temperature unit. The fast drop of the potential energy near $R=R_{vdW}$ can quantum reflect the wavefunction of the Efimov bound state $\psi_1(R)$.}
\label{fig2}
\end{figure}

We begin with quantum scattering of three identical atoms with short-range interactions. As in low-energy scattering of two atoms, low-energy scattering of three atoms only affects the lowest partial wave ($s-$wave) where all hyperangular momenta are in their ground states. The lowest partial wave $\psi$ only depends on the hyperradius $R$ and satisfies an effective Schroedinger equation as

\begin{equation}
[-\frac{\hbar^2}{2m}\frac{d^2}{d R^2}+V_{m}(R)+V_{E}(R)]\psi(R)=E\psi(R),
\label{eq2}
\end{equation}

\noindent where $V_m(R)$ is the short-range molecular potential, $\psi(R)$ is the three-body wavefunction with energy $E$. Under Born-Oppenheimer approximation, the hyperspherical radius $R=<r_{12}>=<r_{23}>=<r_{13}>$ equals to the mean separation between any two atoms. The emergence of the first Efimov resonance is associated with the appearance of the first bound state at $E=0$ when the Efimov potential extends to $R=|a_{-}^{(1)}|$, see Fig~\ref{fig2}.

The two potential terms, molecular potential $V_m$ and Efimov potential $V_E$, are drastically different in their energy and length scales. The Efimov potential is long-range and dominates for $R>R_{vdW}$, where $R_{vdW}$ is the three-body van der Waals length. Near $R=R_{vdW}$, where Efimov potential is still about $\sim$mK in temperature unit, the van der Waals potential quickly takes over. At shorter range $R<R_{vdW}$, the molecular potential is much deeper, typically $10^3\sim10^4$~K, and is very complicated due to non-additive terms \cite{Soldan2003tbn} and contributions from channels with different spins and hyperangular momenta.

To estimate the three-body van der Waals length $R_{vdW}$, we note that the asymptotic behavior of the three-body potential can be approximated by the sum of three pair-wise potentials \cite{Standard1993}. Together with the Born-Oppenheimer approximation $R=r_{12}=r_{23}=r_{13}$, the long-range behavior of the molecular potential can be written as

\begin{eqnarray}
V_m(R, \Omega)&\approx& V_{12}(R)+V_{23}(R)+ V_{13}(R)+O(R^{-9}) \nonumber \\
      &\approx& - 3C_6R^{-6}+O(R^{-8}),
\label{eq3}
\end{eqnarray}

\noindent where $C_6$ is the two-body van der Waals coefficient and $\Omega$ denotes all hyperangular variables. Leading order corrections include the quadrupole-quadrupole interaction of the order $\sim R^{-8}$ and the non-additive three-body interaction of the order $R^{-9}$ \cite{Moszynski1995}. Under these approximations, the three-body van der Waals length is related to the two-body length as $R_{vdW}=3^{1/4}r_{vdw}$.

The crossover of the two potentials $V_E(R)$ and $V_m(R)$ near $R=R_{vdW}$ can lead to an interesting quantum reflection effect. Quantum reflections occur when the external force $f=|dV/dx|$ is comparable or greater than the local energy-to-length ratio of the particle $(E-V)k$, where $k$ is the local scattering wavenumber. For an Efimov bound state at $E=0$, this condition evaluated at $R=R_{vdW}$ gives $f/k|V|\approx6$, where the factor $6$ comes from the exponent of the van der Waals potential. This result justifies a strong quantum reflection \cite{reflection}. With the assumption of a full and elastic quantum reflection, the Efimov wavefunction would develop a node near $R_{vdW}$, and the wavefunction amplitude is mostly outside the node, see Fig.~\ref{fig2}.

To accurately determine the position of the node, we note that the nodal position of a wavefunction at low energy is related to the scattering length, which for a van der Waals potential, takes a nominal value of the mean scattering length $\bar{A}=4\pi\Gamma(1/4)^{-2} R_{vdW}$ \cite{Gribakin1993csl,Braaten2006uif}, where $\Gamma(.)$ is the Gamma function. Written in terms of the two-body van der Waals length, the nodal position is given by

\begin{equation}
\bar{A}=4\pi\Gamma(1/4)^{-2} 3^{1/4} r_{vdW}.
\label{eq4}
\end{equation}

Taking $\psi(R=\bar{A})=0$ as a boundary condition and ignoring the short-range potential $V_{m}(R>\bar{A})=0$, we can analytically solve the Schroedinger equation Eq.~\ref{eq2} with the Efimov potential Eq.~\ref{eq1}. The positions of the Efimov resonances can be determined by solving the scattering lengths that extend the Efimov potential far enough to support an eigenstate with zero energy $E=0$. The result for the position of the $N-$th Efimov resonance $a=a_{-}^{(N)}$ is

\begin{equation}
a_-^{(N)}=\exp(\frac{N\pi-\tan^{-1}2s_0}{s_0}) \bar{A}.
\label{eq5}
\end{equation}

This is the main result of the paper. We note that it satisfies the recursion relationship $a_{-}^{(N+1)}=e^{\pi/s_0} a_{-}^{(N)}$, and the ratio of the first Efimov resonance to the two-body van der Waals length is thus given by

\begin{equation}
C=\frac{a_-^{(1)}}{r_{vdW}}=4\pi\Gamma(1/4)^{-2} 3^{1/4} \exp(\frac{\pi-\tan^{-1}2s_0}{s_0}).
\label{eq6}
\end{equation}

\noindent For identical particles with $s\approx1.00624$, we obtain $C=9.48...$, which agrees well with the majority of the measurements.

\begin{figure}\includegraphics[width=3.5 in]{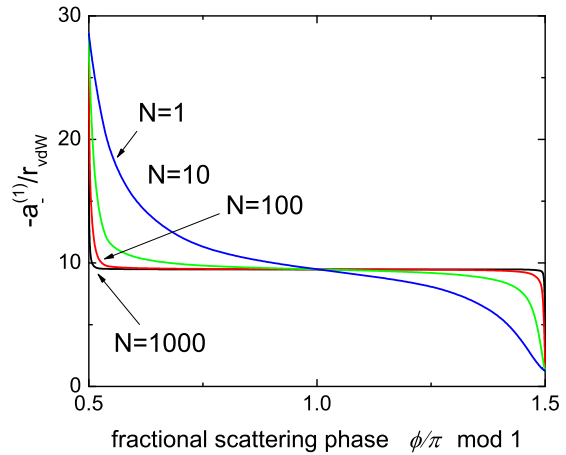}\\
\caption{Dependence of Efimov resonance position on the short-range scattering phase shift. Based on a square well potential model with the radius $\bar{A}$ given by Eq.~\ref{eq4}, the first Efimov resonance position is calculated for identical particles ($s_0\approx1.00624$) in the presence of $N=1, 10, 100$ and 1000 short-range vibrational bound states. The scattering phase shift varies in the range of $\phi=(N-1/2)\pi\sim (N+1/2)\pi$ for $N$ bound states. The result is shown as a function of ($\phi/\pi$ mod 1) for convenient comparison.}
\label{fig3}
\end{figure}

In order to gain further understanding why the Efimov resonance position is insensitive to the short-range three-body phase shift, and to test our assumption of strong quantum reflection, we introduce a square well potential model to simulate the short range potential. We assume $V_m(R>\bar{A})=0$ and $V_m(R<\bar{A})=-D$, where $\bar{A}$ sets the length scale of quantum reflection, and the potential depth $D$ determines the short range scattering phase shift as $\phi=\sqrt{2mD}\bar{A}/\hbar$ for a zero energy state. The depth of the potential can also be parameterized by the number of vibrational bound states as $N=[\phi/\pi+1/2]$, where $[x]$ is the integer part of $x$ \cite{Gribakin1993csl}. We can then calculate the location of the first Efimov resonance $a_{-}^{(1)}$ as a function of the phase shift $\phi$ in the vicinity of the potential depths that support $N=1, 10, 100$ and 1000 short-range molecular states, see Fig.~\ref{fig3}.

Our calculation suggests that in a deep potential, $a_{-}^{(1)}/r_{vdw}$ is in general a universal constant and is insensitive to the scattering phase shift $\phi$. This is an expected result since quantum reflection works better when the short range potential is deeper. Near $\phi=(N-1/2)\pi$, where $N=1,2,..$, however, the universality breaks down. Remarkably, the condition $\phi=(N-1/2)\pi$ is exactly that for having a new bound state at the threshold. We can thus conclude that the universal value of $C=-a_{-}^{(1)}/r_{vdw}=const.$ holds as long as the short-range potential is deep enough to support many bound states, and none of them is near the dissociation threshold. For alkali atoms with typically $100$ or more molecular states, our model suggests that even with a randomly-distributed three-body phase shift $\phi$, the probability to obtain $C=9\sim 10$ is $95\%$ or higher. This result explains the universality of the ratio $C$ in many experiments, and the exceptional case of $^{39}$K can potentially be associated with the existence of a bound state near the continuum, which resonantly shifts the Efimov resonance position.

To conclude, we obtained a simple physics picture based on quantum reflection to understand the universal scaling of the Efimov resonance position by the van der Waals length. Quantum reflection of the Efimov wavefunction occurs because of the presence of a very deep and short range molecular potential. Assuming complete quantum reflection near the mean scattering length, we obtain the universal ratio of $C=a_{-}^{(1)}/r_{vdW}=9.48..$, which agrees well with the universal value observed in experiments. Based on a square well model, we show that Efimov resonance positions are insensitive to the three-body phase shift as long as the short range potential supports many bound states, and none of the states are accidentally located near the dissociation continuum. The universality of the Efimov resonance position can potentially be extended to three atoms with unequal masses when two or three pair-wise interactions are resonantly enhanced. Analysis on the role of different van der Waals length scales and scattering phase shifts between different atoms is needed in order to determine the universality of the Efimov energy structure in systems with unequal masses.

We want to emphasize that the universality of Efimov physics in cold atoms remains hypothetical. Our quantum reflection model provides a simple picture to understand the fixed nodal position of the Efimov wavefunction. However, detailed calculations incorporating foreign scattering channels, especially those inducing the Feshbach coupling, as well as non-Born-Oppenheimer, and higher-order potential terms, are needed to test the genuineness of the universality, as well as to understand the appearance of the non-universal case, such as $^{39}$K.

We thank C. Greene, B. Esry, S. Jochim, S. Kokkelmans and R. Grimm for discussions, and S. Jochim for careful reading of the manuscript. This work is supported by NSF Award No. PHY-0747907 and AFOSR-MURI cold molecules grant.


\begin{thebibliography}{99}
\bibitem{reflection}
For simplicity, we only discuss the equilateral triangle configuration. Quantum reflection can also occur and potentially be stronger with other hyperangular configurations \cite{yujun}.
\end{thebibliography}
\end{document}